\documentclass[reqno]{amsart}
\usepackage{graphicx,graphics}
\usepackage{subfig}

\usepackage{booktabs}
\usepackage{multirow}

\theoremstyle{definition}

\numberwithin{thm}{section}
\numberwithin{lem}{section}
\numberwithin{coll}{section}
\numberwithin{rem}{section}
\numberwithin{exm}{section}
\numberwithin{prop}{section}
\numberwithin{equation}{section}

\usepackage{mathtools}

\numberwithin{equation}{section}

\begin{document}

\centerline {\textsc{\large Stock Prices as Janardan Galton Watson Process }}
\vspace{0.5in}

\begin{center}
%Author 1 with Address\\
% \vspace{.1in}
   Ali Saeb \footnote{Corresponding author: ali.saeb@gmail.com}\\
   Department of Economic Sciences,\\
   Indian Institute of Science Education and Research,\\
   Bhopal 462 066, India

\end{center}

\vspace{1in}

%#############################################################################
%                                  ABSTRACT                                  #
%#############################################################################
%\begin{abstract}

\noindent {\bf Abstract:}
Janardan (1980) introduces a class of offspring distributions that sandwich between Bernoulli and Poisson. This paper extends the Janardan Galton Watson (JGW) branching process as a model of stock prices. In this article, the return value over time $t$ depends on the initial close price, which shows the number of offspring, has a role in the expectation of return and probability of extinction after the passage at time $t.$ Suppose the number of offspring in $t^{th}$ generation is zero, (i.e., called extinction of model at time $t$) is equivalent with negative return values over time $[0, t].$ We also introduce the Algorithm that detecting the trend of stock markets. 
\vspace{0.5in}

\vspace{0.2in} \noindent {\bf Keywords:} Branching Process, Galton Watson Process, Janardan model.

\vspace{0.5in}

\vspace{0.2in} \noindent {\bf AMS subject classification:} 60J80, 60J85

\newpage
\section{Introduction}
The Galton-Watson process is a Markov chain $\{X_n: n= 0,1,2,\ldots\}$ on the nonnegative integers with offspring distribution $\{p_k\}.$ The process is also called Bienayame-Galton-Watson (BGW) process. The interest lies mainly in the probability distribution of $X_n$ and the probability that $X_n\to 0$ for some $n,$ i.e., the probability of ultimate extinction of the family. By extinction of the process it is meant that the random sequence $\{X_n\}$ consists of zeroes for all except a finite number of values of $n.$ In order words, extinction occurs when $\Pr(X_n=0)=1$ for some value of $n.$ Karlin and Taylor (1981) and Ramiga (1977) investigated the behavior for offspring distributions, such as Poisson, Binomial, Geometric, and uniform families.
An application of the branching process, Epps (1996) studies the BGW branching process is proposed to model the short-term behavior of stock prices. The four models are chosen Bernoulli, Poisson, Translated log series and geometric to illustrate the random generation model. It obtains estimates of the probability of eventual failure of the firms in the sample and the expected time until failure. Mitov et al. (2009) examine the pricing of barrier options when the price of the underlying asset is modeled by a branching process in a random environment. 

With similar GWP and inspired by nature, Mitchell (1975) experimentally investigated the oviposition behavior of bruchid beetles on mung beans. 
Janardan (1980) developed a stochastic model, assuming a rate $\lambda$ until one egg is laid and a rate $\eta( < \lambda)$ on a bean having already one egg.  The beetles are selective in laying eggs on mung beans, in the sense that the chance of laying a second egg on a bean already having one egg is smaller than the chance of laying an egg on a bean having no egg. Janardan et al. (1995,  1997) show that the probability mass function is the specialty of this class of distributions in that it gives a family of distributions sandwiched between Bernoulli and Poisson. They also developed a new stochastic model incorporating this random phenomenon. %Vasudeva and Saeb (2013) study the asymptotically behavior of Janardan Galton Watson (JGW). 
 
In this article, our main interest is to investigate the JGW process in stock closing price. In the next section, we study the properties of the random generation number of JGW. We also study the probability of extinction and expected return over $[0, t]$ and give our main Algorithm which using in the stock markets in section 3. %The results are shown in section 3.
 Appendices A and B containing the graphs and tables.

Throughout the manuscript, we shall denote the derivative by $(\cdot)'.$ 
%Also, we employ the notation, $\floor{\cdot}$ is the floor.

\section{Janardan Model}
Let $\xi$ denote the offspring random variable (rv) with the probability generating function (pgf) $\Pi(s)=\sum_{i=0}^{\infty}{p_i s^{i}},\;\; 0\leq s \leq 1,$
where $p_i=\Pr(\xi=i),\;\;i\geq 0.$ The size of the $n^{th}$ generation is the sum of the total offspring of the individuals of the previous generation. That is, $X_n=\sum_{i=1}^{X_{n-1}}\xi_i,$ where, $X_n$ denotes the number of children born to the $n^{th}$ person in the $(n-1)^{th}$ generation. The members of $X_0$ are called the ancestors.\\
%With no loss generality $X_0=1.$
Let the pgf of $X_n$ be denoted by $\Pi_n(s)=\Pi_{n-1}(\Pi(s)),$ $0\leq s\leq 1.$ Some of the characteristics of study are the size of the $n^{th}$ generation, eventual extinction or explosion of the species and so on. The pgf of $X_n$ can be used to find the mean, variance and higher moments of $X_n.$ If we let $\mu=\Pi'(1)$ be the
mean number  of offspring, then this recurrence relation entails
%We observe that
\begin{eqnarray}\label{e1}
        E(X_n)=\Pi'_n(1)=(\Pi'(1))^n=\mu^n,
\end{eqnarray}
Also, $\sigma^2$ is the variance of $\xi,$ then the variance of $X_n$ is obtained,
\begin{eqnarray}\label{e2}
Var(X_n)= \left\lbrace	
	\begin{array}{l l}
	\frac{\mu^{n}(\mu^n-1)}{\mu(\mu-1)}\sigma^2, &\;\text{if}\;\; \mu\neq 1,\\
	&\\
	n\sigma^2, &\;\text{if}\;\; \mu=1. \\
 \end{array}
  \right.					
\end{eqnarray}
%\[Var(X_n)=\Pi''(1)+\Pi'(1)-(\Pi'(1))^2.\]
Given a branching process, the probability of eventual extinction is the smallest positive root of the equation $s = \Pi(s).$ If $\mu\leq 1,$ that is, in the subcritical and critical cases, the extinction probability is equal to $1.$ 

Let $\Pr(X(t)= n)$ be the probability there are $n$ offspring at
time $t.$ Now, we consider a Janardan Galton Watson (JGW) process, with probability mass function (pmf)
	\begin{eqnarray} 
		\Pr(X(t)=0)&=&e^{-t\lambda},\nonumber\\
		\Pr(X(t)=1)&=&\dfrac{\lambda}{\eta-\lambda}(e^{-t\lambda}-e^{-t\eta}),\nonumber\\
		\vdots &&\nonumber\\
		\Pr(X(t)=m)&=&\dfrac{\lambda \eta^{m-1}}	{(\eta-\lambda)^m}\left(e^{-t\lambda}-e^{-t\eta}\sum_{j=0}^{m-1}\dfrac{(\eta-\lambda)^j}{j!}\right),\nonumber
	\end{eqnarray}
	$m\geq 2,\;\; \lambda>0,$ and $0<\eta<\lambda$ with the pgf of $X(t)$ is given by
	\begin{equation} \label{e2.1}
		\Pi(s,t;\eta,\lambda)=\dfrac{1}{\eta(s-1)+\lambda}\left((s-1)(\eta-\lambda)e^{-t\lambda}+s\lambda e^{-t\eta(1-s)}\right),\;\; 0\leq s\leq 1.
	\end{equation}
	Since, $X(1)$ is the number of members of first generation, it is straightforward to show for first generation $X(1)$, 
	\begin{eqnarray}
		\mu=E(X(1))=E(\xi)=\frac{\eta}{\lambda}(e^{-\lambda}+\lambda-1)+(1-e^{-\lambda}),\;\; 0<\eta<\lambda.\label{e2.5}
	\end{eqnarray}
	Similarly, the variance of the distribution is given by 
\begin{eqnarray}
\sigma^2=\eta^2-\mu^2
+(1-e^{-\lambda})\left(1-\frac{\eta}{\lambda}\right)\left(1-\frac{2\eta}{\lambda}\right)+\eta\left(3-\frac{2\eta}{\lambda}\right).\label{e2.6}
\end{eqnarray}
Now, we get
$\lim_{\eta\rightarrow\lambda} \Pi(s;\eta,\lambda)=e^{-\lambda(1-s)}$ and $\lim_{\eta\rightarrow 0}\Pi(s;\eta,\lambda)=e^{-\lambda}+s(1-e^{-\lambda}),$ which the pgf of Poisson with parameter $\lambda$ and Bernoulli with parameter $(1-e^{-\lambda})$ respectively. 
It trivially notes that $E\xi$ is an increasing function of $\eta$ with
\[\lim_{\eta\to 0}E\xi=(1-e^{-\lambda})\qquad \text{and} \qquad \lim_{\eta \to\lambda}E\xi=\lambda.\]
Vasudeva and Saeb (2013) show that, if $E\xi\leq 1$ then, for all $\lambda<1,$ and $\eta>g(\lambda),$ (or, for all $\lambda\geq 1,$ and $\eta<g(\lambda)$) where, $g(\lambda)=\dfrac{\lambda e^{-\lambda}}{e^{-\lambda}-(1-\lambda)}$ the branching process is sub-critical; it is  Critical, if $\eta=g(\lambda).$

The branching process is super-critical if $E\xi>1.$ In other hand, the probability of extinction is the fractional root of $\Pi(s;\eta,\lambda)=s$ when $\lambda<1,$ $\eta<g(\lambda)$ (or, $\lambda>1,$ $\eta>g(\lambda)$). 

{\bf Estimation of Parameters:} The method of maximum likelihood estimations, one can see that solving for $\lambda$ and $\eta$ is highly complicated. We hence adopt the following method of repeated moment estimation. We know that the offspring rv $\xi$ takes the integer positive values. Let $f_0, f_1, \ldots$ are the class frequencies based on a random sample of size $n.$ Define a new rv $Z$ with
	\begin{eqnarray*}
Z= \left\lbrace	
	\begin{array}{l l}
	0, &\;\text{if}\;\; \xi\geq 1,\\
	1, &\;\text{if}\;\; \xi=0; \\
 \end{array}
  \right.					
\end{eqnarray*}
i.e. $Z=1$ if there are no offspring and $=0$ if there are one or more offspring. $Z$ is a Bernoulli rv with $\Pr(Z=1)=e^{-\lambda}$ and $\Pr(Z=0)=1-e^{-\lambda}.$ Thus $EZ=e^{-\lambda}.$ Hence $\lambda$ is estimated by moment estimation, i.e. $EZ=\frac{\sum_{j=1}^{\infty}Z_j}{n}=\frac{f_0}{n},$ where $f_0$ is the total relative frequency of zeroes. The solution is given by 
\begin{eqnarray}
\hat{\lambda}=\log n-\log f_0.\label{e6}
\end{eqnarray}
Having obtained $\hat{\lambda},$ $\hat{\eta}$ is obtained by equating the first moment $E\xi$ with the sample mean $\bar{X}=\frac{\sum_{j=1}^{\infty}jf_j}{n}$ and substituting $\hat{\lambda}$ for $\lambda.$ We get
\begin{eqnarray}
\hat{\eta}=\frac{\hat{\lambda}(\bar{X}-1+e^{-\hat{\lambda}})}{e^{-\hat{\lambda}}+\hat{\lambda}-1}.\label{e7}
\end{eqnarray}
Such a technique has been suggested in Anscombe (1950) and the same is also used in Janardan (1980).

\section{Illustration To Stock Price}
Let $X_0$ represent the price of one share of stock at time $t=0.$ At time $t>0$ the price is $X(t)=X_{N(t)},$ where $N(t)\geq 0$ represents the random number of generations during $[0,t].$ The Poisson process $\{N(t)\},$ with $N_0=0,$ and parameter $t\theta$ is assumed to have stationary, independent increments and to be independent of $X_n$ and $\{\xi_i\}.$ Henceforth, we refer to $P(t)=\frac{X(t)}{X_0}
$ as the random generation process. The $P(t)$ is equivalent the number of offspring $\xi$ at each time $t.$ The pgf of the process $P(t)$ is given by
\begin{eqnarray}
\phi(t,s)=E(s^{P_{N(t)}})=EE(s^{P(t)})=\sum_{n=0}^{\infty}\frac{(\theta t)^n}{n!}e^{-\theta t}\Pi_n(s),\label{3.1}
\end{eqnarray}
where, $E(s^{P(t)})=\Pi_n(s).$ From (\ref{e1}) and (\ref{e2}) and some calculations we derive the following
formulas for the mean and the variance of the process $P(t)$ (see details, eq. 2.12, Epps, 1996),
\begin{eqnarray}
E(P(t))&=&\phi'_s(t,1)=E(\Pi'_{N(t)}(1))=E\mu^{N(t)},\nonumber\\
&=& e^{\theta t(\mu-1)}.\label{EXt}
\end{eqnarray}
The variance of $P(t)$ is given by
\begin{eqnarray}\label{VXt}
Var(P(t))= \left\lbrace	
	\begin{array}{l l}
	\left(\frac{\sigma^2}{X_0\;\mu(\mu-1)}\right)(e^{\theta t(\mu^2-1)}-e^{\theta t(\mu-1)})+\left(e^{\theta t(\mu^2-1)}-e^{2\theta t(\mu-1)}\right), &\text{if } \mu\neq 1,\\
	\frac{\sigma^2\theta t}{X_0}, &\text{if } \mu=1,\\
 \end{array}
  \right.					
\end{eqnarray}
where, $\sigma^2=Var(\xi).$

%	\begin{eqnarray}
%E(R(t))= \left\lbrace	
%	\begin{array}{l l}
%	 e^{\theta t(\mu-1)}-1\label{ERT}, &\;\text{if}\;\; \mu\neq 1,\\
%	0, &\;\text{if}\;\;\mu=1 \\
% \end{array}
%  \right.					
%\end{eqnarray}
%where, $\mu=E\xi.$

%When the branching process is super-critical,
The next interesting measure is the probability of extinction.
%Since, extinction occurs with probability $1,$
Let $P(t)$ is the size of generation at time $t$, $\Pr(P(t)=0)$ is the probability of failure at time $t$ a proper rv which is equivalent $\Pr(X(t)< X_0).$ From (\ref{3.1})  we have 
\begin{eqnarray}
		\Pr(P(t)=0)=\phi(t,0)=\sum_{n=0}^{\infty}\Pi_n(0)\Pr(N(t)=n),\label{prob}
\end{eqnarray}
where $\Pi_n(0)=\Pi_{n-1}(\Pi(0)).$ 

Now, we get an estimate of $\theta,$ $\hat{\theta},$ is then obtained from $\hat{\lambda}.$ Suppose that the $\{\xi_i\}$ are positive integer valued random variables with a common distribution function $F.$ Refering Feller (1966) and Theorem A, Janardan et al. (1995), define 
$S_{N(t)}=0$ if $N(t)=0,$ and $=\sum_{i=1}^{N(t)}\xi_i$ if $N(t)\geq 1.$ Then $S_{N(t)}$ is Poisson if and only if $F$ is Bernoulli. if $S_{N(t)}$ is compound Poisson distribution  with parameter $\lambda$ and $N(t)$ is Poisson with parameter $t\theta,$ then it may be easily observed that $t\theta\geq \lambda$ necessarily. Further, the parameter of Bernoulli distribution is $\lambda/t\theta$. In particular if $\lambda=\theta,$ then $\xi_1,\xi_2,\ldots$ are degenerate at $1.$ Recalling that $E(N(t))=E(S_{N(t)})$; i.e., expected number of generations until failure is the mean number of generations per unit time. Then $\Pr(\xi=0)=1$ which means that  
 the probability $S_N=0$ given that $N=0$ is degenerated one.

\newpage
In what follows, we introduce the algorithm for estimating the parameters and further calculations. 

\begin{figure}[!ht]
  \centering
   \subfloat{\includegraphics[width=.85\textwidth]{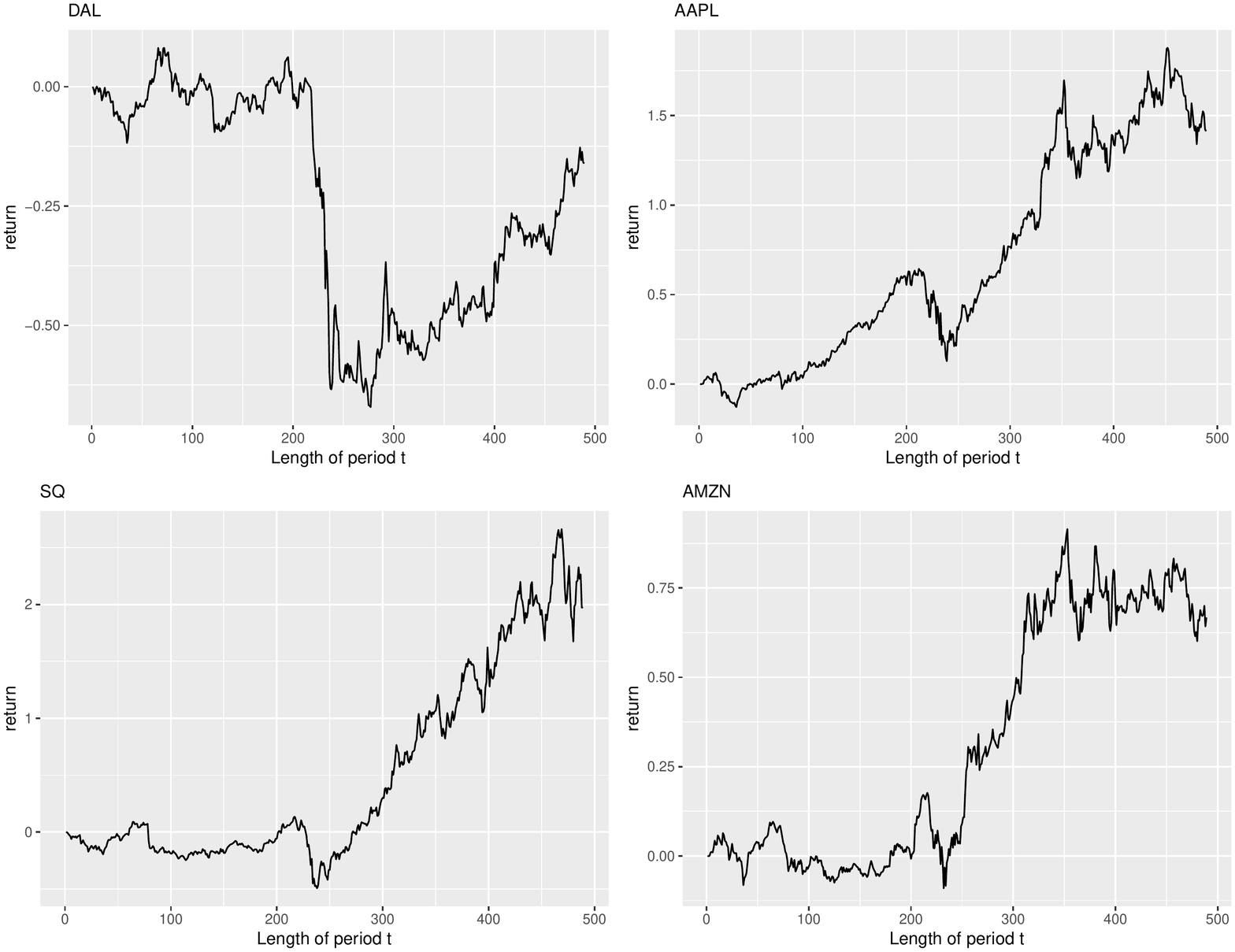}} \label{Graph}
    \caption{\small{the total return of assets over two years}}\label{G1}
    \subfloat{\includegraphics[width=.85\textwidth]{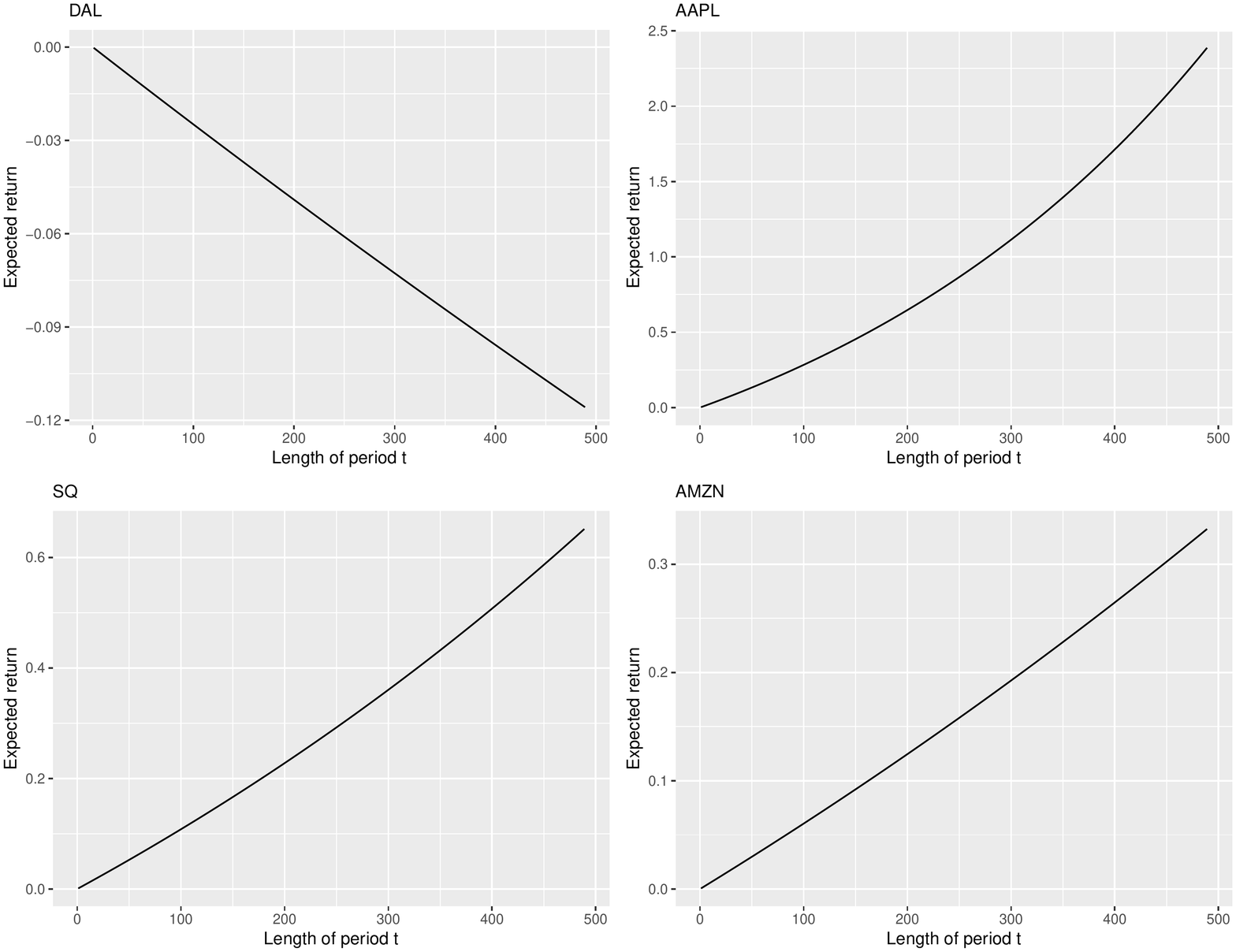}} \label{Graph}
    \caption{\small{Expected return over the time}}\label{G2}
\end{figure}

\begin{figure}
  \centering
   \subfloat{\includegraphics[width=.9\textwidth]{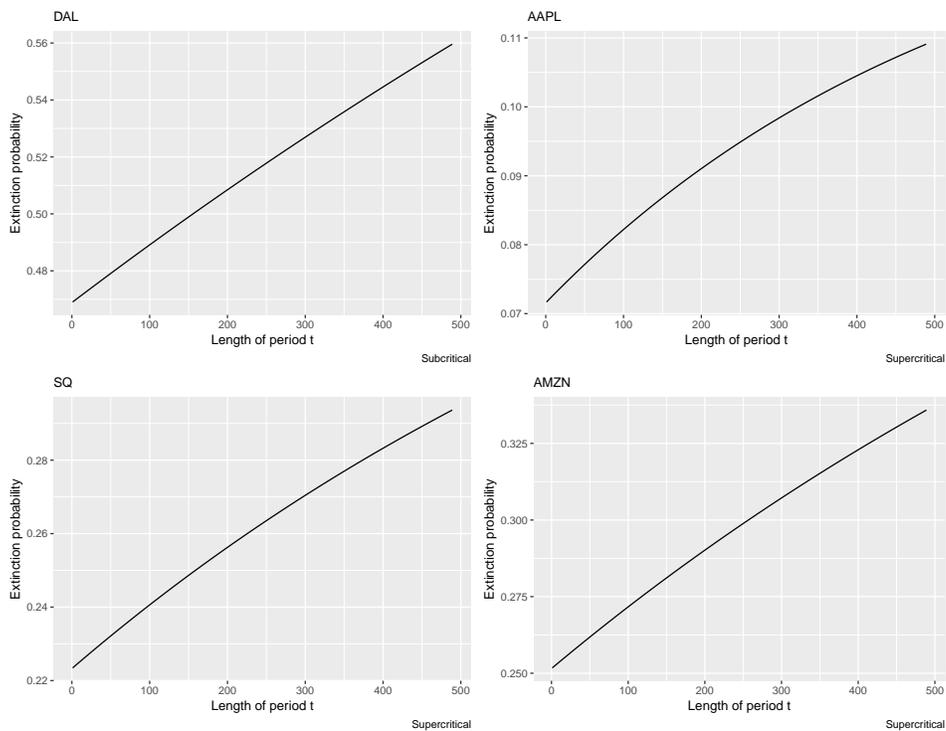}} \label{Graph}
    \caption{\small{Probability of extinction in different time}}\label{G3}
\end{figure}

{\bf Algorithm:}
\begin{enumerate}
\item Calculate the daily returns over $[0,t]$ are $r_1,r_2,\ldots, r_{n}$, where, $r_i=\frac{X_i}{X_{i-1}}-1$. Let $\{\xi_i\}$ is the cumulative ratio of stock price. It obtains with 
$$\xi_i=\frac{X_i}{X_0}=\Pi_{i=1}^{n}(1+r_i).$$ 
Note that, $\xi$ is the number of offspring in the first generation.
%is given by $P_i(1)=\floor{\xi_i}.$ 
\item Count the number of $\xi$ is less than $1$ and call it as $f_0.$
\item Compute the value of mean from the $\{\xi_i\}$ and call it as $\bar{x}.$
\item Estimate $\lambda$ and $\eta$ from \ref{e6} and \ref{e7} and find $g(\hat\lambda).$
\item Consider the time of $t.$ Estimate the parameter $\theta,$ with $\hat\theta=\lambda/t.$
\item Substitute the values of $\hat\lambda,$ $\hat\eta$ and $\hat\theta$ in equations \ref{e2.5}, \ref{e2.6} and find the values of mean and variance of generations in \ref{EXt} and \ref{VXt}.
\item Check the status of extinction. 
%If $\hat\lambda<1$ the JGW is sub-critical. The JGW is sub-critical, critical or super-critical according as $\hat\eta<g(\hat\lambda),\;\hat\eta=g(\hat\lambda),\;\hat\eta>g(\hat\lambda).$ 
\item If JGW is the super-critical we are calculated the probability of extinction by using \ref{prob}.
\end{enumerate}
Now, we study the random generation process on the real data set of the NYSE. Firstly, the parameters of JGW-Poisson estimate for the NYSE with the following four stocks: DAL, AAPL, SQ, and AMZN. We selected two years the stock price from April-2019 to April-2021. The 490 observations correspond closely to the number of trading days in two years. Graph. \ref{G1} shows the total return $P(t)-1 =(X_t-X_0)/X_0$ over two years period, where, $X_t$ is the market close price at time $t.$ The proportion price in first generation given by the Janardan model with parameters $\lambda$, $\eta$ and $N(t)$ is the random number in the first generation over period time $[0,t]$ follows the Poisson with parameter $t\hat\theta.$ Brief results for estimates of sample mean and variance of offspring in time $[0, 500]$ corresponding with $\mu$, $\sigma^2$ and estimated the parameters $\lambda$, $\eta$, $\theta$, and $g$ are shown in table \ref{T1} for each offspring model for the four tickers. We estimate $\hat{\lambda}$ by using the observed proportion of negative total return. Substituting $\hat\lambda$ in (\ref{e7}) given $\hat\eta.$ If $\hat\eta>\hat\lambda$ then we estimate $\hat\lambda$ with sample mean of observation, $\mu.$ In this scenario, we obtained the Poisson-Possion model. To measure the performance of the model, we show root mean square error (RMSE) in table \ref{T1}. % From (\ref{EXt}), this is defined as follows  
%\[RMSE=\sqrt{\frac{\sum_{t=1}^n\left(P(t)-E(P(t))\right)^2}{n}},\]
% Here, we compute the number of offspring with proportion of daily closing price at each assets i.e, $\xi_t=X_t/X_{t-1}.$ The mean and variance of $\xi$ are shown in Table \ref{T1}. $\tilde{P}(t)$ can be represented  as $\Pi_{i=1}^{i} \xi_i=R(t)+1$ and $P(t)=\floor{\tilde{P(t)}}.$  
The values of graph \ref{G2} shows the expected return on different values of time, (i.e, $E(P(t))-1$). In this graph, the ticker of DAL is subcritical which shows that the expected total return is decreasing and these values are increasing for the other supercritical models. Graph \ref{G3} presented the probability of extension over time $[0, 500].$ For example, the probability of extinction of DAL is $60\%$ at time $500.$ The other tickers have a small probability of extinction at the same time. The values of the estimated expectation, variance, probability of extinction model and actual value for some specific time are presented in table \ref{T2}. 

\newpage
%##############################################################
%                                  APPENDIX                                  #
%############################################################
\appendix
\section{Graphs}

\newpage

\newpage
\section{Tables}
\begin{table}[!ht]
\begin{tabular}{lllllllll}
\toprule
Tickers & $\mu$ &   $\sigma^2$ &   $\hat{\lambda}$ & $\hat{\eta}$ & $\hat{\theta}$ &  RMSE & $g$& Status\\
\toprule
  DAL & $0.7575$ & $0.7575$ & $0.7575$ & $0.7575$ & $0.001$ & $0.2784$ & $1.5692$
 & Subcritical\\
  AAPL & $1.6907$ & $1.0607$ & $2.637$ & $1.1765$ & $0.0036$ & $0.3348$ & $0.1105$
 & Supercritical\\
 SQ & $1.4994$ & $1.4994$ & $1.4994$ & $1.4994$ & $0.0021$ & $0.7121$ & $0.4632$ & Supercritical\\
  AMZN & $1.3104$ & $1.2248$ & $1.3802$ & $1.2278$ & $0.0019$ & $0.2959$ & $0.5496$
 & Supercritical\\
\bottomrule
\end{tabular}
 \caption{\small{Estimated the parameters of JGW-Poisson}}\label{T1}
\end{table}

\begin{table}[!ht]
\begin{tabular}{clllll}
\toprule
\multicolumn{1}{l}{Tickers} & working days (t) & $E(P(t))$ & $Var(P(t))$ & $\Pr(P(t)=0)$ & Actual value \\
\multicolumn{1}{l}{} &  &  &  & & $P(t)$\\ \midrule
\multicolumn{1}{l}{\multirow{4}{*}{DAL}} & \multicolumn{1}{|l}{90}   & 0.9776 & 0.0067 & 0.4871 & 0.9966\\
\multicolumn{1}{l}{}     &
\multicolumn{1}{|l}{180}   &  0.9557 & 0.0128 & 0.5047 & 1.02\\
\multicolumn{1}{l}{}     &
\multicolumn{1}{|l}{365}  & 0.9122 & 0.0239 & 0.5385 & 0.5107\\
\multicolumn{1}{l}{}     &
\multicolumn{1}{|l}{489}  & 0.8842 & 0.0303 & 0.5596 & 0.8392  \\
\midrule
\multicolumn{1}{l}{\multirow{4}{*}{AAPL}} & \multicolumn{1}{|l}{90}   & 1.2518 & 0.2673 & 0.0812 & 1.0573 \\
\multicolumn{1}{l}{}     &
\multicolumn{1}{|l}{180}   & 1.5669 &  0.9065 & 0.0894 & 1.4572 \\
\multicolumn{1}{l}{}     &
\multicolumn{1}{|l}{365}  &  2.4861 & 5.4818 & 0.1025 & 2.2132 \\
\multicolumn{1}{l}{}     &
\multicolumn{1}{|l}{489}  &   3.3875 & 15.3542 & 0.1091 & 2.4125  \\
\midrule
\multicolumn{1}{l}{\multirow{4}{*}{SQ}} & \multicolumn{1}{|l}{90}   & 1.0967 &  0.0582 & 0.2389 & 0.8487\\
\multicolumn{1}{l}{}     &
\multicolumn{1}{|l}{180}   &   1.2028 & 0.1431 & 0.2533 & 0.8481 \\
\multicolumn{1}{l}{}     &
\multicolumn{1}{|l}{365}  &  1.4541 &  0.4445 & 0.2789 & 2.0016 \\
\multicolumn{1}{l}{}     &
\multicolumn{1}{|l}{489}  &  1.6513 & 0.7927 & 0.2936 & 2.978 \\
\midrule
\multicolumn{1}{l}{\multirow{4}{*}{AMZN}} & \multicolumn{1}{|l}{90}   &   1.0542 &  0.0185 & 0.2697 & 0.9848 \\
\multicolumn{1}{l}{}     &
\multicolumn{1}{|l}{180}   &  1.1114  & 0.0413 & 0.2866 & 1.0134    \\
\multicolumn{1}{l}{}     &
\multicolumn{1}{|l}{365}  & 1.2389  & 0.1059 & 0.3176 & 1.6054\\
\multicolumn{1}{l}{}     &
\multicolumn{1}{|l}{489}  &  1.3324 & 0.1661 & 0.3359 & 1.6675\\
\bottomrule
\end{tabular}
 \caption{\small{Estimated the expectation, variance, probability of extinction and actual value in some particular working days}}\label{T2}
\end{table}

\end{document}